\documentclass[preprint,aps,floats,nofootinbib,amssymb]{revtex4}
\pdfoutput=1 
\usepackage[T1]{fontenc} 
\usepackage{hyperref}
\usepackage[utf8]{inputenc}
\usepackage{amssymb}
\usepackage{latexsym}
\usepackage{graphics}
\usepackage{graphicx}
\usepackage{feynmf}
\usepackage{color}
\usepackage[sort&compress]{natbib}
\usepackage{epsf,epsfig}
\usepackage{amsmath}
\usepackage{hyperref}
\usepackage{subfigure}
\usepackage{mathrsfs}

\newcommand{\be}{\begin{equation*}}
\newcommand{\ee}{\end{equation*}}
\newcommand{\ba}{\begin{eqnarray*}}
\newcommand{\ea}{\end{eqnarray*}}

\newcommand{\bw}{\begin{widetext}}
\newcommand{\ew}{\end{widetext}}

\begin{document}         
\title{\vspace*{1.in}\large
Seeding baryonic dark matter}
\vspace*{0.5cm}

\author{Dom\`enec Espriu\footnote{espriu@icc.ub.edu}}
\affiliation{
Departament de F\'isica Qu\`antica i Astrof\'isica and
Institut de Ci\`encies del Cosmos (ICCUB), \\
Universitat de Barcelona, 
Mart\'i Franqu\`es 1, 08028 Barcelona, Spain}

\begin{abstract}
  It has been proposed that primordial quark pellets or PQPs ---ultra-dense quark‑matter mini-stars--- formed at $T\sim  1$ GeV
  maybe a good candidate accounting for the dark matter of the universe. If correct, a large component of dark matter would consist of very
  compact objects with a maximum mass of 10$^{-2}$ $M_\odot$ and radii of approximately 100 m, although smaller objects
  would be much more abundant and encompass the bulk of the dark matter content.
  Here we describe a seemingly viable formation mechanism based on the enhancement of the local baryon density when
  supra-horizon Peccei–Quinn domain walls formed at an earlier epoch sweep and accumulate quarks and gluons before entering
  the horizon. Assuming an efficient baryon concentration by contracting Peccei–Quinn domain walls, we derive the resulting primordial
  quark pellet mass spectrum, minimum mass and cosmological abundance
  
\end{abstract}

\maketitle


\section{Introduction}
Finding a suitable candidate for the dark mass of the universe is proving to be very challenging\cite{dark}. An interesting
cosmic coincidence is that the current baryon to photon ratio $\sim 10^{-10}$ \cite{Btogamma}
roughly agrees with the ratio between dark matter and  radiation energy densities in the primordial universe when
$T=1 $ GeV. This ratio stays constant in earlier times if the temperature just mentioned marks the dividing line where 
dark matter transitions from being relativistic to non-relativistic. Some people interpret this coincidence as an indication
that dark matter is of baryonic nature.

There have been many attempts to translate this idea into workable models. The so-called MACHOs\cite{machos} and
strange matter nuggets\cite{nuggets} are two possibilities considered in the past. Astronomical observations are
however quite constraining
and it is usually accepted that objects heavier than $10^{-8}-10^{-7}$ solar masses could constitute at most a few per cent
of the gravitationally observed dark mass, whereas lighter objects may evade these
constrains so far.\cite{Niikura2019}-\cite{OGLEnew} 

  Recently\cite{espriu} it was demonstrated that primordial quark pellets
or PQPs---ultra-dense quark‑matter mini-stars formed
  at $T\sim  1$ GeV ---would naturally arise in a radiation dominated early universe if enough rare baryon overdensities
  could be efficiently produced.
  Solving the Tolman–Oppenheimer–Volkoff (TOV) equation with a hot quark equation of state,
 stable solutions with a maximum mass of 10$^{-2}$ $M_\odot$ and radii of approximately 100 m could be found, although
 the formation mechanism favoured much lighter objects that would become dominant and
 could provide a large fraction of the dark matter. The present proposal originated from a
 study \cite{KRX} that advocated primordial neutron stars as possible dark matter candidates. The PQP proposal
 entirely avoids the need for a matter dominated primordial universe, necessary in \cite{KRX}, and it appears
 more conservative and natural.

 Once formed, PQPs cool and evolve into either
 mini-neutron stars or stable strange-matter nuggets, depending on the QCD ground state.
 This mechanism could  reproduce the present
 dark‑matter density without altering Big Bang Nucleosynthesis\cite{Fields2020} or requiring
an  entropy injection. PQPs could evade microlensing constraints if their mass is below 10$^{-8} \ M_\odot$.
  PQPs thus potentially constitute a conservative, Standard‑Model‑based, and observationally
  testable candidate to the dark‑matter puzzle.  

  In \cite{espriu} it was suggested that the natural  mechanism giving rise to the baryon overdensities
  required for seeding
  the PQPs is the coalescence of domain walls produced after the Peccei-Quinn (PQ) transition\cite{PQ}-\cite{other}.
  As bubbles re-enter the horizon they sweep and collect most baryons in their way\cite{sweep,Liu2019,shellard}
  and this is the origin of the overdensities that would eventually give rise to the PQPs.
  Assuming an efficient baryon concentration by contracting Peccei–Quinn domain walls, we derive the resulting
  primordial quark pellet mass spectrum, the minimum mass and the cosmological abundance
  
  It is important to clarify that this mechanism does not rely on the standard QCD axion potential, which only becomes relevant
  at the QCD scale. PQ domain walls formed before or during inflation naturally stretch to supra‑horizon scales and survive until
  horizon re‑entry, provided reheating does not restore the PQ symmetry \cite{Dineetal}.
  The domain walls required in our scenario could be present in a variety of
  well-motivated cosmological scenarios.\cite{Liu2019,sweep,earlydomains}. In Section VI
  a microphysical tentative model is presented.

  The cosmological analysis presented in Sections II–V assumes only that a periodic axion potential exists at temperatures
  in the TeV region, producing a network of domain walls. The pausible microphysics that generates this potential
  is presented in Section VI, but the cosmological results do not depend
  on the specific realization.

  The present study has to be read in conjunction with the results of \cite{espriu}. While \cite{espriu} described the
  existence, equilibrium, stability, maximum mass and observational limits of PQPs, the present manuscript
  establishes their formation, minimum mass, mass spectrum and abundance.

  The purpose of this work is to establish the cosmological plausibility of this mechanism rather than to provide a
  complete microscopic description of domain wall formation and baryon capture. We show that, under physically
  motivated assumptions about the domain-wall population and its interaction with the primordial plasma, the
  relevant scales are mutually consistent: the primordial baryon asymmetry can be concentrated into compact
  quark matter objects with masses extending into the range required for dark matter, while the resulting abundance can
  reproduce the observed dark-matter density. The microscopic reflection and transmission of quarks at the wall,
  the detailed nonlinear evolution of the wall network, and a complete ultraviolet realization remain open questions.
  We regard these as quantitative problems for future work rather than as assumptions that alter the basic
  cosmological scaling presented here.

\section{Primordial Quark Pellets: the formation mechanism}

One prior assumption is the presence of a Peccei-Quinn transition at some high energy scale in the
early universe. We are considering QCD axions, whereby the relation $m_a^2 f_a^2 = \textrm{constant}$ is assumed
to hold\cite{QCDaxion}. Current estimated indicate that the axion constant $f_a$ could be somewhere in the
range (in GeV) $10^9 \le f_a \le 10^{12}$. 

The formation mechanism does require that this transition takes place before or in an inflationary period so that a
typical PQ domain would eventually engulf many horizon patches at some initial temperature $T_i$ that will be
later specified. From  $T=T_i$ onwards the conventional cosmic flow is assumed to take place.

We shall also assume that this mechanism is totally independent of the origin of the baryon asymmetry. In fact
we need to accept that a net baryonic charge of the right amount exists at $T_i$. This is not a prediction
of this model.

The domain walls will form bubbles with typical sizes $R_i$. They will likely follow a scale-free
distribution of the form\cite{shellard}
\begin{equation}\label{eq:scaleinv}
  \frac{dn}{dR_i}\sim \frac{1}{R_i^4}
\end{equation}
Inflation is expected to blow up the bubbles and eventually extending over $N$ causal patches at $T_i$.
Any bubbles  not extending beyond a single causal patch are likely to quickly collapse due to their surface tension
and play no subsequent role whatsoever.

Our initial condition is thus well defined:  a scale-free distribution of bubbles at temperature $T=T_i$,
formed by thick walls that prevent the free passage of baryons. As the universe
cools and expands, the bubbles expand with the cosmic flow at a rate $R_i\sim 1/T$. They cannot interact
yet because their walls
are causally disconnected. On the other hand, the Hubble radius grows
like $1/H\sim T^{-2}$ implying that the bubbles will sooner or later enter the horizon. At that point
the surface tension of the domain wall will
have to compensate the difference in pressures due to the excess baryons inside the bubble
\begin{equation}\label{eq:localbalance}
  \Delta p = \frac{\sigma}{R}.
\end{equation}
Would both pressures be identical, the domain wall would collapse at the speed of light. This
is not the case if the baryon density on both sides is different and a dynamical equilibrium is reached.
The surface tension $\sigma$ should be computable in a microphysics model. As we will see, the contribution
due to the surface tension is always much smaller that the typical pressure even though it plays
a crucial role in the proposed mechanism.

Radiation density decreases as $1/T^4$, whereas non-relativistic matter density dilutes itself as $1/T^3$, implying
that once cold dark matter is formed it eventually dominates. Conversely,
baryons are very rare in the early universe. For instance, at 1 GeV the total energy density
is 20.31 GeV$^4$ but the baryon fraction needed to describe baryonic dark matter and visible baryonic matter
is just $\beta= 5.8\times 10^{-10}$. This fraction stays constant until 
dark matter in the form of PQPs is formed.

In order to create the PQPs, rather large overdensities are needed. As discussed in \cite{espriu}
at the epoch where $T=1$ GeV, to produce an object with a mass 10$^{-8}$ $M_\odot$ per horizon one needs
around 10$^3$ times more baryons than those contained on average in a horizon patch. In other words, while the
total baryon/radiation balance is correct, compact objects could not possibly form unless baryons are lumped together.
The formation and evolution of Peccei-Quinn bubbles constitute a natural mechanism whereby baryons from neighbouring horizon patches
concentrate in a region of a single horizon at sufficiently low temperatures.

For the PQP production mechanism to work as expected, gravity needs to be able to form the pellets
at a temperature between > 200 MeV and 1.27 GeV \cite{espriu}. That means that the density has to be
locally sufficiently large at that epoch. After entering the horizon, the bubbles keep quarks together,
densely packed, decoupling
them from the global thermal bath. As $T$ further decreases, the internal and
external pressures converge, the surface tension contribution becomes very small compared to the internal energy,
and the configuration settles into a TOV solution. Indeed it is easy to check that
for those macroscopic objects $\sigma/R$ is negligible. The final fate of the walls is being radiated away via axion emission.

\subsection{The sweep and collect mechanism}
During the evolution, the condition $f_a \gg T$ has to hold, meaning that the wall, with a thickness
$\Delta \approx 1/m_a$ is almost impenetrable for relativistic baryons at temperature $T$.
By the time the horizon is entered, the bubbles have collected a huge baryon overdensity.

The total baryonic mass collected by one bubble is
\begin{equation}
M = \epsilon \, \beta_i \, N \, M_H(T_i) \label{eq:swept_base}
\end{equation}
where $\epsilon$ is the capture efficiency (fraction of swept baryons retained),
$\beta_i \equiv (\rho_B/\rho_{\rm tot})_{T_i}$ is the baryon fraction at formation,
$N$ is the number of horizon patches engulfed, and $M_H(T_i)$ is
the horizon mass at $T_i$. Everything is highly relativistic at that epoch so $\rho_{\rm tot}=
\rho_{\rm rad}$.

The physical radius of the bubble at formation time is
\begin{equation}
R_i = N^{1/3} H_i^{-1} \sim N^{1/3} \frac{M_{\rm Pl}}{T_i^2},
\end{equation}
where $M_{\rm Pl}$ is the Planck mass and $H_i^{-1} \sim M_{\rm Pl}/T_i^2$ is the horizon size during
radiation domination.
The volume of the bubble is
\begin{equation}
V_i \sim R_i^3 \sim N \frac{M_{\rm Pl}^3}{T_i^6}.
\end{equation}

As the universe expands from $T_i$ to $T_c$, the temperature when dark matter
is formed, the bubble is carried by the Hubble flow. Since it is suprahorizon for most of its existence,
its physical radius scales with the scale factor $a \propto 1/T$
\begin{equation}
R_c = R_i \, \frac{a(T_c)}{a(T_i)} = R_i \, \frac{T_i}{T_c}.
\end{equation}
Substituting $R_i$,
\begin{equation}\label{radius}
R_c \sim N^{1/3} \frac{M_{\rm Pl}}{T_i T_c}. 
\end{equation}
Ideally, if the mechanism would work seamlessly, the total mass of the resulting PQPs would be
very close to the total baryon
energy density collected in $R_c$.

We note that the previous equation approximately matches with the results obtained from the solution
of the TOV equation. If we set $T_i= 1$ TeV and $N=1$ for the smallest pellets we get $\sim 1 $ m, where
the TOV equation gives $\sim 0.1$ m.

\subsection{Causal condition for PQP formation at $T_c$}

For the bubble to contribute to the PQP formation at $T_c$, it must have entered the horizon before that epoch. The horizon size at $T_c$ is:
\begin{equation}
H_c^{-1} \sim \frac{M_{\rm Pl}}{T_c^2}.
\end{equation}
The condition $R_c < H_c^{-1}$ gives
\begin{equation}
N^{1/3} \frac{M_{\rm Pl}}{T_i T_c} < \frac{M_{\rm Pl}}{T_c^2},
\end{equation}
or
\begin{equation}
  T_i > N^{1/3} \, T_c .
  \label{eq:correct_causal}
\end{equation}

The enclosed mass will be
\begin{equation}
M = \epsilon \, \beta_i \, N \, M_H(T_c) \frac{T_c^2}{T_i^2} 
\end{equation}
Substituting the previous causal bound (Eq.~\ref{eq:correct_causal}) into the mass 
and taking the limiting case $T_i \sim N^{1/3} T_c $, we have
\begin{equation}
   M \sim \epsilon \, \beta \, N^{\frac13} M_H(T_c) .
\end{equation}
Note that $\beta_i=\beta_c=\beta$ stays constant as long as dark matter is not present.

The horizon mass at $T_c$ is
\begin{equation}
M_H(T_c)=\frac43 \pi \rho_{\rm rad} H_c^{-3} \sim  \frac{M_{\rm Pl}^3}{T_c^2}.
\end{equation}

In the previous reasoning the number of relativistic degrees of freedom $g_*$ is assumed to stay constant between
$T_i$ and $T_c$ for simplicity. This is not correct but the modifications are small.

\subsection{Minimum and maximum masses}

The smallest possible bubble engulfs exactly one horizon patch. This gives $T_i \gtrsim T_c $,
which is trivially satisfied for any formation temperature above $1\ \text{GeV}$. The absolute minimum mass is
\begin{equation}
M_{\min} \sim \epsilon \beta \, M_H(T_c).
\end{equation}
At $T_c = 1\ \text{GeV}$, the horizon mass is $M_H \simeq 0.062\, M_\odot$. Thus assuming $\epsilon=1$
\begin{equation}
M_{\min} \simeq 5.8 \times10^{-10} \times 0.062\, M_\odot \simeq 3.6\times10^{-11}\, M_\odot.
\end{equation}
This is the absolute lower bound set by causality and baryon conservation.
In physical terms, this is a sphere of radius $\sim 10\ \text{cm}$ containing the mass of a small asteroid.

The TOV integration with the quark equation of state at $T_c=1\ \text{GeV}$ gives the maximal configuration
\cite{espriu}
\begin{equation}
M_{\max} \simeq 0.013\, M_\odot \simeq 1.45 \times 10^{55}\ \text{GeV}, \qquad R_{\max} \simeq 88\ \text{m}.
\end{equation}
This would need the participation of $10^9$ horizons at $T_i$.

\subsection{Capture efficiency}

At this stage we will treat the capture efficiency  $\epsilon$ phenomenologically: the observed
dark-matter-to-baryon ratio
will fix the value required by the scenario to work. 
The capture efficiency is not a free parameter of the formation dynamics, but it should be determined by the
properties of the wall and the kinetic energy of the relativistic baryons. However, we will
do reverse engineering and we will adjust its value assuming that is
cosmologically fixed by baryon number conservation. The argument goes as follows: 
Let the total baryonic mass of the universe be partitioned into two reservoirs:
The fraction $\epsilon$ that is successfully swept up and trapped inside the collapsing domain wall bubbles, eventually forming
the PQP dark matter.

The fraction $1-\epsilon$ that escapes through the moving wall during the sweep remains in the intergalactic medium to
eventually form the visible stars, gas, and baryonic structures we observe today.
Since PQPs are composed entirely of standard model quarks, the total mass in PQPs today is $\Omega_{DM}$ and the
total mass in ordinary visible baryons is $\Omega_B$ Because baryon number is strictly conserved, the ratio of these two
components is exactly equal to the ratio of the trapped to the escaped fractions:
\begin{equation}
  \frac{\Omega_{DM}}{\Omega_B}= \frac{\epsilon}{1-\epsilon}
    \end{equation}
Using the Planck 2018 values\cite{Planck2018} this gives $\epsilon = 0.84 $.
It should be emphasized that this number is an effective parameter required by cosmology rather
than as an independent prediction. The actual value will depend on the scaling regime, bubble distribution
and sweep efficiency.

The present work does not attempt a microscopic calculation of quark reflection or transmission through PQ domain walls.
Instead, it investigates the cosmological implications of efficient baryon concentration,
leaving the detailed particle-level dynamics for future work.
However, the value $\epsilon=0.84$ is very reasonable. If $f_a$ is large enough, assuming a simple square well profile
for the barrier and the WKB approximation, the probability of penetrating the barrier at 1 TeV (see below) will be
$\sim \exp(-f_a/m_a)$, an extremely small number, and it will remain negligible down to the 1 GeV range of
temperatures. The difference between 0.84 and 1  has to be mostly attributed to inefficiencies in the domain wall
recombination trapping the baryons.

We are well aware that the sweep-and-collect mechanism has to be flawless for these numbers to be taken rigorously, but
it is encouraging that the figures are coming out roughly correct.
When the 0.84 factor is included, the minimum mass is reduced to $3 \times 10^{-11}$ solar masses.

\section{Further cosmological considerations}

At this point we need to put several pieces of the
puzzle together. Let us consider first Eq. (\ref{eq:correct_causal}): in order to be able to 
encompass the whole spectrum of masses, we need at least that $T_i > 10^3 T_c$. That is, greater than
or around 1 TeV. As long as the bubble walls do not enter the horizon they are protected, but the
smallest ones (which also are the most abundant) enter the horizon almost immediately below
$T_i$ because $N\simeq 1$. We have to guarantee that the high-density baryon spots created by the
entering bubbles persist in such an
environment.

As previously explained, small bubbles enter first and the last to enter are those that will
eventually contribute to the largest quark pellets allowed by the TOV equation. These are very rare anyway
so it is most important to understand the evolution of the small bubbles. When
they enter the horizon they are very hot; the energy density behaves as $T^4$ multiplied by the
respective numbers of degrees of freedom (inside and outside). Thermal equilibrium
is totally out of the question at this stage and so is forming extended objects. The equilibrium
and compactness of the baryon-rich bubble is enforced by the surface tension; that is, the pressure
difference $\Delta p$ is neutralized by the pressure due to the surface tension.
Thanks to this, they do not completely follow the cosmic flow, being subject to their own dynamical constraints

\subsection{Wall survival and the dissolution depth}

In order to ensure that the topological configurations survive long enough to compress the trapped baryonic plasma rather
than washing out into the ambient thermal bath, we must track the accumulated integrated dissolution
depth $\mathcal{I}$ as the universe expands and cools from an initial temperature $T_i$ to a final temperature $T_c$. 
In a radiation-dominated background, the relation between a cosmic time step $dt$ and a temperature step $dT$ is
\begin{equation}
dt = - \alpha \frac{M_{\text{Pl}}}{T^3} dT
\end{equation}
where $\alpha = \sqrt{90 / (8\pi^3 g_*)}$. The thermal scattering
dissolution rate of the axion wall scales non-linearly with temperature. On dimensional grounds
 $\Gamma_{\text{th}}(T) \simeq \frac{T^3}{f_a^2}$,
The total accumulated dissolution depth $\mathcal{I}$ experienced by the configuration is given by the time integral:
\begin{equation}
  \mathcal{I} = \int_{t(T_i)}^{t(T_c)} \Gamma_{\text{th}}(T) \, dt
  = \int_{T_c}^{T_i} \left( \frac{T^3}{f_a^2} \right) \left( \alpha \frac{M_{\text{Pl}}}{T^3} \right) dT
\end{equation}
Crucially, the $T^3$ power from the thermal interaction rate exactly cancels the $T^3$ volume expansion dilution factor
embedded within $dt$. This leaves a purely linear integrated dependence on the temperature drop
\begin{equation}
\mathcal{I} =  \frac{\alpha  M_{\text{Pl}}}{f_a^2} (T_i - T_c)
\end{equation}
The integral is dominated by the high temperature boundary, reducing simply to
\begin{equation}
\mathcal{I} \approx \frac{\alpha  M_{\text{Pl}} T_i}{f_a^2}
\end{equation}
In order to guarantee wall survival and ensure the capture efficiency $\epsilon$ is not degraded by early thermal melting,
we require $\mathcal{I} < 1$. This maps out a strict upper bound on the allowable maximum temperature $T_i$
for a given axion decay constant $f_a$
\begin{equation}
\label{eq:T_bound}
T_i < \frac{f_a^2}{\alpha  M_{\text{Pl}}}
\end{equation}

If the cosmic history satisfies Eq.~\eqref{eq:T_bound}, the domain walls are protected against thermal fluctuations.
Once this protection condition is satisfied, the configurations transition safely into the low-temperature phase
($T \sim 1 \text{ GeV}$). Before that temperature is reached the contribution from the surface tension becomes negligible.
The quark pellet becomes a stable configuration ---a solution of the TOV equation--- thanks to the equilibrium
between pressures, as described in\cite{espriu}.

Thus, the baryon-rich  bubble content effectively 'wait' until the ambient pressure is low enough to
allow a stable configuration. We regard Eq. (\ref{eq:T_bound}) as quite relevant as it relates the reheating temperature to the
axion structure constant.

\subsection{The wall domination bound}

There are several considerations that need to be taken into account in the framework of the
sweep-and-collect mechanism. The first one is that domain walls should be far from dominating the energy
budget of the universe at any moment.

The surface tension of the PQ domain walls $\sigma = m_a f_a^2$. After entering the horizon the total
domain wall energy has to be
compared to the radiation enclosed in the horizon as radiation; that is
\begin{equation}\label{DWenergy1}
  4\pi H^{-2} \sigma  \ll \rho_{\text rad} \frac43 \pi H^{-3}.
\end{equation}
Up to a factor of order 1, this implies $\sigma \ll M_{\text Pl} T^2$ or in the worst
situation
\begin{equation}\label{DWenergy2}
  f_a^2 \ll \frac{M_{Pl} T_c^2}{m_a}.
\end{equation}
We have to keep this condition in mind when we address the microphysics.

When the domain walls set in, after the PQ breaking took place,  the configuration of the axion field is random
in each domain and not necessarily aligned with the corresponding vacua.
The energy density in bubble $i$ will typically be $\Lambda^4 \theta_i^2$, with
$\Lambda$ being the scale of the new physics that trigger the
appearance of the axion periodic potential and $\theta_i$
of order 1.
A fraction of the misalignment energy is invested in the domain wall network, which will
eventually decay into relativistic
axions and redshifts away, but the residual coherent oscillations will count
as dark matter abundance and it must be checked that no too much cold dark matter 
is produced in this way.

Oscillations start when $H(T) \simeq m_a$. The axion coupling constant $f_a$ enters this condition
when the axion mass is referred to the scale of the new physics $m_a \sim \Lambda^2/f_a$ and
this sets again an upper bound for $f_a$.

These constraints have to be checked in the microphysics model of choice. In addition it has
to be checked that the axion is a QCD axion -i.e. it solves the strong $CP$ problem.
This issue is discussed in Section VI.

\section{The scale-invariant mass function}

A scale-invariant domain wall network in the scaling regime yields a bubble size distribution given
by Equation \ref{eq:scaleinv}.
Since $R \propto N^{1/3}$  and \(M \propto N\), we have
\begin{equation}
 \frac{dn}{dM} = A \, M^{-2} , \label{eq:mass_function}
\end{equation}
where $A$ is a normalization constant.

At $T=1\ \text{GeV}$, the radiation energy density is
\begin{equation}
\rho_{\rm rad} = \frac{\pi^2}{30} g_* T^4, \qquad g_* = 61.75,
\end{equation}
giving \(\rho_{\rm rad} \simeq 20.31\ \text{GeV}^4\), as
was previously quoted.
Therefore
\begin{equation}
\rho_{\rm DM}( 1\ \mathrm{GeV}) \simeq 5.8\times10^{-10} \times 20.31 \simeq 1.18\times10^{-8}\ \text{GeV}^4. \label{eq:rho_DM_T1}
\end{equation}

Integrating the mass function:
\begin{equation}
\rho_{\rm DM}(1\ \mathrm{GeV}) = \int_{M_{\min}}^{M_{\max}} M \, A M^{-2} \, dM = A \ln\left(\frac{M_{\max}}{M_{\min}}\right).
\end{equation}
Hence
\begin{equation}
A = \frac{\rho_{\rm DM}(T_1)}{\ln(M_{\max}/M_{\min})}. \label{eq:A_norm}
\end{equation}
Numerically, \(M_{\max}/M_{\min} \simeq 0.013 / (3\times10^{-11}) \simeq 4\times10^8\), so
\begin{equation}
\ln(M_{\max}/M_{\min}) \simeq 19.8,
\end{equation}
where the 0.84 efficiency in the sweep-and-collect process has been taken into account.
Finally
\begin{equation}
A \simeq \frac{1.18\times10^{-8}}{19.8} \simeq 6 \times10^{-10}\ \text{GeV}^4. \label{eq:A_value}
\end{equation}

A caveat: it has been assumed that PQPs of all masses up to the maximal ones $\sim 10^{-2} \ M_\odot$ can be produced
at 1 GeV. As stated elsewhere, this requires a bubble originally encompassing the baryons
contained in $10^9$ horizons. We somehow regard this as being unlikely and an lower upper limit should probably
be imposed. If this is the case, the mass function
would be even more  skewed towards lighter masses. However, this issue cannot be solved
rigorously without detailed simulations.

\section{Abundance and spacing today}

The local dark matter density is \(\rho_{\rm local} \simeq 0.01\, M_\odot\,\text{pc}^{-3}\). The total number density of PQPs today is:
\begin{equation}
n_{\rm total} = \int_{M_{\min}}^{M_{\max}} A M^{-2} dM = \frac{A}{M_{\min}} \left(1 - \frac{M_{\min}}{M_{\max}}\right) \simeq \frac{A}{M_{\min}}.
\end{equation}
Substituting \(M_{\min} = 3 \times 10^{-11}\,M_\odot\) and \(A = \rho_{\rm DM}/\ln(M_{\max}/M_{\min})\):
\begin{equation}
n_{\rm total} \simeq \frac{0.01\, M_\odot/\text{pc}^3}{M_{\min} \ln(M_{\max}/M_{\min})}
\simeq \frac{0.01}{3.0\times10^{-11} \times 19.8} \simeq 1.7\times10^7\ \text{pc}^{-3}.
\end{equation}

The average separation is:
\begin{equation}
d = n_{\rm total}^{-1/3} \simeq (1.7\times10^7)^{-1/3}\ \text{pc} \simeq 3.9 \times10^{-3}\ \text{pc} \simeq 810\ \text{AU}.
\end{equation}

The masses range from 10 Jupiter-sized objects, corresponding to the larger mass $\sim 10^{-2}\ M_\odot$ to asteroid or meteor size
objects with masses in the $10^{-11}\ M_\odot$ class, with radii ranging from 80 m
down to 10 cm.

The mass function $dn/dM$ implies that each logarithmic decade contributes equally to the total dark matter density.
Table I shows the distribution of mass, number density, and average spacing across the full mass range, from the causal minimum
up to the TOV maximum. The smallest objects are the most numerous, with 
an average spacing of $\sim 838$ AU, while the largest objects are rare, with 
0.15 per cubic parsec and spacings of order a few light-years. This hierarchical structure naturally explains why PQPs have
evaded detection: the dominant population is below the sensitivity threshold of current microlensing surveys, while the larger objects are too rare to be statistically significant.

\begin{table}[h]
\centering
\caption{Distribution of dark matter mass across decades of the PQP mass function. 
The total mass fraction per decade is constant: $\simeq 11\%$. 
Radii are computed from the TOV scaling $R \propto M^{1/3}$, normalized to $R_{\max} = 88$ m at $M_{\max} = 0.013 M_\odot$. 
The average spacing is calculated within each decade.}
\begin{tabular}{|c|c|c|c|}
\hline
\textbf{Mass decade} & \textbf{Typical radius}  & \textbf{Number density} & \textbf{Average spacing} \\
$(M_\odot)$ & $(\text{m})$ & $(\text{pc}^{-3})$ & $(\text{AU})$ \\
\hline
$10^{-11} - 10^{-10}$ & $0.08 - 0.17$ &  $\sim 1.5 \times 10^7$ & 838 \\
$10^{-10} - 10^{-9}$  & $0.17 - 0.37$ &  $\sim 1.5 \times 10^6$ & 1,806 \\
$10^{-9} - 10^{-8}$   & $0.37 - 0.80$ &  $\sim 1.5 \times 10^5$ & 3,888 \\
$10^{-8} - 10^{-7}$   & $0.80 - 1.7$  &  $\sim 1.5 \times 10^4$ & 8,376 \\
$10^{-7} - 10^{-6}$   & $1.7 - 3.7$   &  $\sim 1.5 \times 10^3$ & 18,040 \\
$10^{-6} - 10^{-5}$   & $3.7 - 8.0$   &  $\sim 1.5 \times 10^2$ & 38,860 \\
$10^{-5} - 10^{-4}$   & $8.0 - 17$    &  $\sim 1.5 \times 10^1$ & 83,700 \\
$10^{-4} - 10^{-3}$   & $17 - 37$     &  $\sim 1.5 \times 10^0$ & 180,300 \\
$10^{-3} - 10^{-2}$   & $37 - 80$     &  $\sim 1.5 \times 10^{-1}$ & 388,400 \\
\hline
\end{tabular}
\label{tab:mass_decades}
\end{table}

The overall average spacing across all objects is $\sim 810$ AU, which is nearly identical to the spacing of the smallest decade (838 AU).
This coincidence occurs because the $M^{-2}$ spectrum is steep enough that the smallest decade contains $\sim 90\%$ of all objects;
the larger, rarer objects contribute negligibly to the overall number density and thus barely shift the average distance when all
objects are considered together.

\section{Microphysics of the axion sector}

The formation mechanism described in the preceding sections requires a periodic potential for the axion
field that is present at temperatures well above $\Lambda_{\rm QCD}$. QCD instantons alone cannot provide
this: they only turn on at $T \sim \Lambda_{\rm QCD}$, and the resulting potential is far too weak to
generate the domain walls we require at $T > 1$ TeV. The potential must therefore be generated by new
physics. In this section we describe a realization of this picture and identify the parameter range in
which the model is cosmologically consistent. The reader interested only in the cosmological consequences
may skip this section on a first reading. It should also be stressed that the model discussed below is just a possible one.

We shall assume the existence of a hidden non-Abelian gauge group $G_h$ that confines at a scale
$\Lambda \sim \text{few TeV}$. The axion couples anomalously to both QCD and $G_h$,
\begin{equation}
\mathcal{L} \supset \frac{a}{f_a}\left(c_1\,G\tilde{G} + c_2\,H\tilde{H}\right),
\label{eq:axion-couplings}
\end{equation}
where $G$ and $H$ denote the QCD and hidden field strengths and $c_1$, $c_2$ are anomaly coefficients.
The hidden sector generates the periodic potential
\begin{equation}
V(a) = \chi(T)\left[1 - \cos\!\left(\frac{a}{f_a} + \delta\right)\right],
\label{eq:Vh}
\end{equation}
where $\chi(T)$ is the temperature-dependent topological susceptibility of $G_h$, and $\delta$ is a
relative phase. The zero-temperature susceptibility is $\chi(0) \sim \Lambda^4$, and at high temperatures it is
suppressed by the dilute instanton gas result \cite{dilute}
\begin{equation}
\chi(T) \simeq \Lambda^4 \left(\frac{\Lambda}{T}\right)^{b},
\qquad T \gg \Lambda,
\label{eq:chihighT}
\end{equation}
The value of $b$ is model dependent. In QCD it takes a value $b\simeq 8$, also supported
by lattice calculations. This temperature dependence will be important
for the cosmological evolution described below.

The total axion potential is the sum of the QCD and hidden contributions,
\begin{equation}
V_{\text total}(a) = \chi(T)\left[1 - \cos\!\left(\frac{a}{f_a} + \delta\right)\right]
     + \chi_{\rm QCD}(T)\left[1 - \cos\!\left(\frac{a}{f_a} + \bar\theta\right)\right],
\label{eq:Vtotal}
\end{equation}
where $\bar\theta$ is the physical strong CP phase (including the CKM contribution) and
$\chi_{\rm QCD}(T)$ is the QCD topological susceptibility, which turns on at $T \sim \Lambda_{\rm QCD}$.

If we assume that the dark sector is Higgsed at a scale $v$ such that
$\Lambda_{\rm QCD} \ll v \ll \Lambda$, 
the gauge bosons of $G_h$ acquire masses, instantons cease to exist, and the hidden
susceptibility is exponentially suppressed. The axion potential therefore has a two-stage structure:

\begin{itemize}
\item For $T > v$: both sectors contribute, and since $\chi_h \gg \chi_{\rm QCD}$ the potential is
dominated by the hidden sector. This is the epoch at which the domain walls form.
\item For $T < v$: the hidden contribution switches off, and only QCD remains. The axion becomes the
standard QCD axion, with mass $m_a \simeq \Lambda_{\rm QCD}^2/f_a$, and solves the strong $CP$ problem
in the usual way.
\end{itemize}

For the axion to solve strong $CP$, the minimum of the combined potential must coincide with the QCD
$CP$-conserving point. Since the hidden sector dominates at high $T$, the axion sits at
$a/f_a \simeq -\delta$. Requiring this to coincide with the QCD minimum at $a/f_a = -\bar\theta$ gives
$ \delta = \bar\theta$.
This is not automatic: it would require a symmetry relating the two sectors. Several 
mechanisms could be invoked: a mirror symmetry between the hidden and visible sectors, $Z_N$ alignment,
forcing the minima to coincide, or the existence of a  GUT  encompassing both sectors.
However, having a large $CP$ violating sector at high energies above the TeV scale could be
relevant.

We stress that the alignment condition only needs to hold at
high $T$ if $CP$ violation needs to be suppressed at that epoch. Below $v$, the hidden potential switches off,
and the axion relaxes to the QCD minimum
regardless of the initial value of $\delta$.

Let us now examine whether the microphysics we just sketched fulfill the expected conditions

\subsection{Axion contribution to dark matter}

The axion starts oscillating when its mass becomes comparable to the Hubble rate,
\begin{equation}\label{oscill}
m_a(T_{\rm osc}) = H(T_{\rm osc}),
\qquad
m_a(T) = \frac{\sqrt{\chi(T)}}{f_a}.
\end{equation}
Using Eq.~\eqref{eq:chihighT}, this gives
\begin{equation}
T_{\rm osc} \simeq \Lambda
\left(\frac{M_{\text Pl}}{1.66 \sqrt{g_*} f_a}\right)^{\frac{2}{b+4}}.
\label{eq:Tosc}
\end{equation}
For $\Lambda= 3$~TeV, $f_a = 10^{12}$~GeV, $g_* = 100$, $b = 8$:
\begin{equation}
T_{\rm osc} \simeq 30~\text{TeV} \simeq 10\,\Lambda.
\end{equation}
The axion is therefore oscillating well before the hidden sector confines, in the regime where the
dilute instanton gas approximation is reliable.

The energy density at oscillation is
\begin{equation}
\rho_a(T_{\rm osc}) = \chi(T_{\rm osc})\,\theta_i^2,
\end{equation}
where $\theta_i$ is the initial misalignment angle. The ratio to the radiation energy density is
\begin{equation}
\frac{\rho_a}{\rho_{\rm rad}}\bigg|_{T_{\rm osc}}
= \frac{30}{\pi^2 g_*}\left(\frac{\Lambda}{T_{\rm osc}}\right)^{4+b} \theta_i^2,
\end{equation}
which grows as $T^{-1}$ as the universe cools. Taking into account the modification of the
degrees of freedom, this result has to be  multiplied by $\sim 10^{17}$ to get the present
densities. For $f_a \sim 10^{12}$~GeV and $\Lambda \sim$~TeV the model is compatible with the
observed dark matter density for $\theta_i \sim 1$. No fine tuning of the initial alignment is needed.

The functional form of Eq.~\eqref{eq:chihighT} is the dilute instanton gas
result \cite{dilute}. The exponent $b$ is model-dependent and, in
principle, requires knowledge of the matter content and gauge group of
the hidden sector. It turns out, however, that the axion abundance is
only weakly sensitive to $b$. To see this, note that combining
Eqs.~\eqref{eq:chihighT} and~\eqref{eq:Tosc} gives
\begin{equation}
\left(\frac{\Lambda}{T_{\rm osc}}\right)^{b+4}
=
\frac{1.66\sqrt{g_*}\,f_a}{M_{\rm Pl}},
\end{equation}
so the factor $(\Lambda/T_{\rm osc})^{b+4}$ entering the energy density
is $b$-independent. The residual $b$-dependence enters only through
$T_{\rm osc}$ in the redshift factor $a_0/a_{\rm osc}\propto T_{\rm osc}$,
which varies by at most a factor of two across the plausible range of
$b$. We therefore take $b$ as a fixed benchmark value and do not pursue
its precise determination further.

A relevant fraction of dark matter originating from axion oscillations is possible, complementing
the contribution of the PQPs.

\subsection{Walls never dominate.}

The domain wall network must not dominate the energy density of
the universe before it decays. This requires the fulfillment of
of Eqs. \ref{DWenergy1} and \ref{DWenergy2}. At energies in the TeV
region Eq. \ref{oscill} reads
\begin{equation}
  m_a\simeq \frac{\Lambda^2}{f_a}
\end{equation}
implying, if we assume $\Lambda\sim 3$ TeV 
\begin{equation}
f_a \lesssim \frac{M_{\rm Pl}\,T^2}{\Lambda^2}
\simeq 10^{19}~\text{GeV}
\left(\frac{\Lambda}{3~\text{TeV}}\right)^{-2}.
\label{eq:wallbound}
\end{equation}
The bound becomes easier to satisfy as $T\to T_c$ and $\Lambda$
is replaced by $\Lambda_{\text QCD}$.

\subsection{The surface tension}

It has been stated in Sections II and III that the contribution from the domain wall to
equilibrate the interior and exterior pressures is small. This is an important point
as the temperature approaches $T_c$, the temperature where the EOS and the chemical
composition allows for the formation of the PQPs. It is of particular
concern for the small pellets as the surface pressure is proportional to $1/R$.

The surface tension is $\sigma \simeq \Lambda^2 f_a$. The radius of the bubble
is given by Eq. \ref{radius}. We have already seen that, up to factors of order unity,
\begin{equation}
  \sigma \ll \rho_{\text rad} H^{-1}.
  \end{equation}
  Therefore
  \begin{equation}
    \frac{\sigma}{R} \ll \rho_{\text rad} \frac{1}{N^{1/3}} \frac{T_i}{T}.
    \end{equation}
  The worst possible situation would correspond to $N=1$ and $T=T_i$ and then
  \begin{equation}
    \frac{\sigma}{R} \ll \rho_{\text rad} ,
    \end{equation}
Q.E.D.

In fact as $T \to T_c \simeq 1 $ GeV, $\Lambda$ becomes $\Lambda_{\text QCD}$ and
the string tension becomes very small. The energy stored in the domainwall
has been progressively radiated away in the form of axions.  This is coherent
with the fact that in the TOV derivation no contribution from the domain wall pressure
was taken into account at all.

\subsection{Other constraints}

It has already been argued why the proposed microphysics model describes
a QCD axion that solves the strong $CP$ problem. We have also seen
that the axion dark matter is not overproduced. The misalignment abundance does not exceed the observed dark
matter density, $\Omega_a h^2 \lesssim 0.12$. Even though a detailed
numerical analysis is needed, this constrains $f_a$ from above, as seen. 
Axions  do not spoil BBN either as they are quickly redshifted away.

It remains to justify the $1/M^2$ distribution that stems from a $1/R^4$ distribution
in suprahorizon domain wall bubbles.

This distribution is the standard result for a topological defect network in the scaling regime,
where self-similarity caused by the rapid expansion implies a scale-invariant size distribution with the exponent fixed by the requirement
that the network energy density is dominated by the horizon scale \cite{shellard,scalingindex}.
The distribution is cut off below the horizon scale, which in our case is implemented by the causal bound $N\ge 1$
of Section II.B. Numerical simulations give exponents in the range $3.5 - 4.5$ quite compatible with
the assumption.

The window
\begin{equation}
10^9~\text{GeV} \lesssim f_a \lesssim 10^{13}~\text{GeV},
\qquad
\Lambda \sim \text{few TeV},
\qquad
\Lambda_{\rm QCD} \ll v \ll \Lambda_h
\end{equation}
is therefore cosmologically viable and seems to lead to a consistent formation scenario.

\section{Conclusions}

We have presented an apparently viable mechanism to seed baryonic dark matter.
Within the proposed framework, a complete, self-consistent mechanism for cold dark matter composed of
primordial quark pellets emerges. 
The mass spectrum is entirely fixed by the scale invariance of the domain wall network and the TOV
stability limit. The lower cutoff is set by causality: $N=1$ yields $M_{\min} \sim 10^{-11}\,M_\odot$.
The observed dark matter abundance normalizes the mass function without any free parameters other
than a  natural value for the capture efficiency $\epsilon \sim 0.84$.
The predominant objects are basketball-sized nuggets with mountain-like masses, spaced $\sim 850$ AU apart,
making them completely invisible to current microlensing surveys and collisionless with Earth, but
the proposal actually describes a continuum of masses that has to be made compatible with the observations.

The observational aspects of the model have not been discussed in detail here but the interested reader can find
additional information in
\cite{espriu}. It would be worth to reconsider existing constraints, future observational opportunities,
possible distinguishing signatures relative to primordial black holes or other macroscopic dark matter candidates.

The seeding mechanism rests on five physical pillars: (i) suprahorizon PQ bubbles from inflation, with
$f_a$ larger than the reheating temperature $T_i$,
(ii) the standard scaling of domain wall networks, (iii) baryon sweeping via axion wall coalescence, (iv)
wall survival via surface tension until TOV solutions can take over, and (v) baryon conservation fixing
the dark matter-to-baryon ratio. All are motivated in the literature or quantitatively
derived here.

We have also described a microphysics model, and argued that it fulfills all the necessary
requirements derived from cosmology. We have emphasized that it is just one of probably several
possible theoretical constructions.

Several ingredients of the scenario remain to be established quantitatively. In particular, a
microscopic calculation of quark reflection and transmission at the PQ wall is needed to determine
the capture efficiency rather than treating it phenomenologically. The detailed evolution and survival
of the wall network, including its behavior across the transition in the axion potential, also merits
numerical study. Finally, the assumed scaling distribution of the initial bubbles should ultimately be
tested in simulations of the relevant network dynamics. None of these questions changes the basic
order-of-magnitude consistency demonstrated here: the baryon content available in the primordial plasma,
the concentration factors associated with rare large wall regions, the resulting compact object mass
scale, and the abundance required for dark matter can all be mutually consistent. The present work should
therefore be viewed as a proof of concept for a new cosmological formation mechanism,
rather than as a complete microscopic realization. Certainly much work is needed before it could
be established.

As a summary, below we quote some of the relevant quantities predicted by the model
\begin{table}[h]
\centering
\begin{tabular}{|l|c|}
\hline
\textbf{Quantity} & \textbf{Value} \\
\hline
Minimum mass ($N=1$) & $\sim 3.0\times10^{-11}\, M_\odot$ \\
Maximum mass (TOV limit) & $0.013\, M_\odot$ \\
Mass function slope & $dn/dM \propto M^{-2}$ \\
Normalization $A$ & $6.0\times10^{-10}\ \text{GeV}^4$ \\
Number density today & $\sim 1.7\times10^7\ \text{pc}^{-3}$ \\
Average spacing & $\sim 810\ \text{AU}$ \\
Capture efficiency required & $\epsilon \sim 0.84$ \\
Reheating temperature & $f_a^2/M_{Pl} > T_i > 1 \text{TeV}$\\
\hline
\end{tabular}
\end{table}

\section{Acknowledgements}
 The financial support from
the State Agency for Research of the Spanish Ministry of Science, Innovation and Universities through the “Unit of Excellence Maria de Maeztu 2025-2028”
award to the Institute of Cosmos Sciences (CEX2024-001451-M) and through project PID2022-136224NB-C21 is acknowledged. We also acknowledge
the support of the Catalan Government through grant 2021-SGR-249.


\begin{thebibliography}{99}

\bibitem{dark}
  G.~Bertone and D.~Hooper, ``A history of dark matter'', \textit{Review of Modern Physics}, \textbf{90}, 45002 (2018).

\bibitem{Btogamma}
  S.~Nussinov, ``Technocosmology: Could a technibaryon excess provide a 'natural' missing mass candidate?'',\textit{Physics Letters B}, \textbf{165}, 55 (1985);
  H.M.~Hodges, ``Mirror baryons as the dark matter", \textit{Physical  Review D}. \textbf{47}, 456 (1993);
  D.B.~Kaplan, M.A.~Luty and K.M.~Zurek, ``Asymmetric dark matter'', \textit{Physical Review D}, \textbf{79},115016 (2009);
  D. Brzeminski and A,Hook, ``Dynamical Explanation of the Dark Matter and Baryon Energy Density Coincidence",
  \textit{Physical  Review Letters}, \textbf{132}, 201001 (2024).

\bibitem{machos}
  B.~Carr, “Baryonic Dark Matter,” \textit{Annual Review Astronomy and Astrophysics}, \textbf{32}, 531 (1994).

\bibitem{nuggets}
  E.~Witten,
  ``Cosmic separation of phases",\textit{Physical Review D}, \textbf{30}, 272 (1984);
  E.~Farhi and R.L.~Jaffe, ``Strange matter," \textit{Physical Review D}, \textbf{30}, 2379 (1984);
 Y.~Bai, A.J.~Long and S.Lu, ``Dark quark nuggets,", \textit{ Physical Review D}, \textbf{99}, 055047 (2019);
 D.M.~Jacobs, G.D.~Starkman and B.W.~Lynn, ``Macro Dark Matter,"
 \textit{Monthly Notices of the Royal Astronomical Society}, \textbf{450}, 3418 (2015).
  
\bibitem{Niikura2019}
H.~Niikura \textit{et al.} (HSC Collaboration),
``Microlensing constraints on primordial black holes with Subaru/Hyper Suprime-Cam'',
\textit{Nature Astron.} \textbf{3}, 524--534 (2019).

\bibitem{Niikura2019ogle}
H.~Niikura \textit{et al.} (OGLE Collaboration),
``Constraints on Earth-mass primordial black holes from OGLE 5-year microlensing events,''
\textit{Phys. Rev. D} \textbf{99}, 083503 (2019).

\bibitem{Tisserand2007}
  P.~Tisserand et al. (EROS-2 Collaboration),
  ``Limits on the MACHO content of the Galactic Halo from the EROS-2 Survey of the Magellanic Clouds''
  \textit{Astronomy \& Astrophysics}, \textbf{469}, 387 (2007).

\bibitem{Witt1994}
H.J.~Witt and S.~Mao, ``Can Finite Source Sizes Be Resolved in Gravitational Microlensing Events?''
  \textit{The Astrophysical Journal}, \textbf{430}, 505 (1994).


\bibitem{Gould1997}          
  A.~Gould and C.~Gaucherel, ``Finite Source Effects in Microlensing Events''
  \textit{Astrophysical Journal}, \textbf{477}, 580 (1997).

\bibitem{OGLEnew}
  P.~Mroz et al.,``No massive black holes in the Milky Way halo'', \textit{Nature}, \textbf{632}, 749 (2024);
  ``Limits on Planetary-mass Primordial Black Holes from the OGLE High-cadence Survey of the Magellanic Clouds'',
  \textit{The Astrophysical Journal Letters}, \textbf{976}, L19 (2024).
  

\bibitem{espriu}
D.~Espriu, ``Do primordial quark pellets solve the dark matter puzzle?'', arXiv:2607.10672
  
 \bibitem{KRX}
  G.~Krnjaic, D.~Rocha and H.~Xiao,
  ``Primordial neutron stars'',
  https://arxiv.org/abs/2604.08651.

\bibitem{Fields2020}
B.~D.~Fields \textit{et al.},
``Big-Bang Nucleosynthesis after Planck,''
\textit{JCAP} \textbf{03}, 010 (2020).

\bibitem{PQ}
  R.~D.~Peccei and H.R.Quinn, ``CP Conservation in the Presence of Pseudoparticles", \textit{Phys. Rev. Lett.}, \textbf{38}, 1440 (1977)

\bibitem{shellard}
  A.~Vilenkin and E.P.S.~Shellard, ``Cosmic strings and other topological defects'', Cambridge University Press (1994);
  I.~Baldes, Y. Gouttenoire and F. Sala, ``Bubbletrons", arXiv:2306.15555 (2023).

\bibitem{Liu2019}
  L. Liu, Z.-K. Guo and R.-G. Cai, ``Primordial Black Holes from Cosmic Domain Walls", arXiv:1901.07072 (2019).

\bibitem{other}  
K.~Harigaya, K.~Kawasaki, M.~Mukaida and T.T.~Yanagida, ``More axions from diluted domain walls", \textit{Phys. Rev. D}, \textbf{94}, 063506 (2016);
M.~Gorghetto, E.~Hardy and G.~Villadoro, ``Early vs late string networks from a minimal QCD Axion", \textit{JHEP,}\textbf{06}, 151 (2018).

\bibitem{sweep}
  A.G.~Cohen, D.B.~Kaplan and A.E.~Nelson, ``Progress in electroweak baryogenesis'', \textit{Annual Review of Nuclear and Particle Science}.\textbf{43}, 27 (1993);
  G.~Dvali, H.~Liu, and T.~Vachaspati, ``Sweeping away the monopole problem'', \textit{Physical  Review  Letters}, \textbf{80}, 2281 (1998);
  T.~Vachaspati, ``Symmetries within domain walls'', \textit{Physical Review D}, \textbf{67}, 125002 (2003);
  Y.~Bai and T.K.~Chen, ``Baryoid Dark Matter from $Z_N$ Domain Walls'', arXiv:2605.13958 (2026)

\bibitem{Dineetal}
  M.~Dine, W.~Fischler, M.~Srednicki, “A simple solution to the strong CP problem with a harmless axion,” \textit{Physics Letters B}, \textbf{104}, 199 (1981);
  J.~Preskill, M.B.~Wise and F.~Wilczek, “Cosmology of the invisible axion,” \textit{Physics Letters B}, \textbf{120}, 127 (1983);
  P.~Sikivie, “Of Axions, Domain Walls and the Early Universe,” \textit{Physical Review Letters}, \textbf{48}, 1156 (1982).

\bibitem{earlydomains}
  K.-F. Lyu and Y. Zhao, "QCD Axion Domain Walls from Super-Cooling First Order Phase Transition," arXiv:2506.19918;
  M. Kawasaki, E. Sonomoto, "Domain wall and isocurvature perturbation problems in a supersymmetric axion model,"
  \textit{Physical Review D}, \textbf{96}, 103502 (2017) ;
  K. Harigaya, K. Kawasaki, M. Mukaida, and T. T. Yanagida, "More axions from diluted domain walls," \textit{Physical Review D},
 \textbf{94}, 063506 (2016);
  S. Nakagawa, Y. Nakai, Y.-C. Qiu, L. Wang, and Y. Wang, "High Reheating Temperature without Axion Domain Walls," arXiv:2509.24812.

  
\bibitem{QCDaxion}
S.~Weinberg, ``A new light boson?'', \textit{Physical review Letters}, \textbf{40}, 223 (1978)


\bibitem{Planck2018}
N.~Aghanim et al. (Planck Collaboration),``Planck 2018 results. VI. Cosmological parameters,"  \textit{Astronomy \& Astrophysics}, \textbf{641}, A6 (2020). 


\bibitem{dilute}
  D.J.~Gross, R.D.~Pisarski and L.G.~Yaffe, ``QCD and instantons at finite temperature''.
  \textit{Review of Modern Physics}, \textbf{53}, 43 (1981); N.~Carabba and E.~Meggiolaro,
  ``Study of some local and global U(1) axial condensates
in QCD at finite temperature'',
  \textit{Physical Review D}, \textbf{105}, 5 (2022) 

\bibitem{scalingindex}
  W.H.~Press, B.S.~Ryden and D.N.~Spergel, "Dynamical evolution of domain walls in an expanding universe,"
  \textit{Astrophysical Journal},\textbf{347}, 590 (1989);
  M.~Hindmarsh, "Analytic scaling solutions for cosmic domain walls," \textit{Physical Review Letters}, \textbf{77}, 4495 (1996)
  


\end{thebibliography}
\end{document}